\documentclass[aps,prd,reprint,nofootinbib,groupedaddress]{revtex4-2}
\usepackage{amsmath, amsthm, amssymb}
\usepackage[svgnames]{xcolor}

\usepackage{accents}
\usepackage{graphics}  %% for \scalebox{}

\definecolor{britishracinggreen}{rgb}{0.0, 0.26, 0.15}
\definecolor{britishracingred}{rgb}{0.23, 0.0, 0.13}
\usepackage{hyperref}
\hypersetup{
    colorlinks=true,
    linkcolor=DarkBlue,
    citecolor=Navy,%britishracingred,%DarkGreen,
    urlcolor=DarkBlue,
    }

\newcommand{\beq}[1]{\begin{equation} \label{#1}}
\newcommand{\eeq}{\end{equation}}

\newcommand{\bea}[1]{\begin{eqnarray} \label{#1}}
\newcommand{\eea}{\end{eqnarray}}

\newcommand{\HIDE}[1]{}

\definecolor{repColor}{rgb}{0.0, 0.0, 0.35} %% Color for spatial quantities
\definecolor{gColor}{rgb}{0.20, 0.0, 0.35} %% Color for gauge quantities
\definecolor{sColor}{RGB}{0, 70, 0} %% Color for spatial quantities
\definecolor{mygray}{gray}{0.2}
\definecolor{igray}{rgb}{0.1,0.1,0.15}

\DeclareRobustCommand{\ColorGauge}[1]{\color{gColor}#1\normalcolor}  %Gauge Coloring
\DeclareRobustCommand{\ColorRepAux}[1]{\color{repColor}#1\normalcolor}  %Parameterization Coloring

\newcommand{\NBT}[1]{\emph{#1}}

  \newcommand{\tskip}{{\hspace{0.8pt}}}   % tiny skip

\newcommand{\LiB}{\Big(}
\newcommand{\RiB}{\Big)}

\newcommand{\Ddim}{\mathrm{d}} %{D}%{{\color{DarkBlue}\mathbf{d}}}  % dimension of space-time
\newcommand{\tx}{t}
\newcommand{\sx}{{\color{sColor}{x}}}

\newcommand{\dsx}{d\sx} %{{d^{^{\,\Ddim{-}1}}\!\!\!\!\sx}}

\newcommand{\sVol}{{\color{sColor}V}}
\newcommand{\sint}{{\mathop{\color{sColor}\textstyle{\int}}}}

\newcommand{\ialt}{alt}

\newcommand{\GR}{GR}
\newcommand{\UMG}{UMG}
\newcommand{\GUMG}{GUMG}

\newcommand{\g}{\gamma}
\newcommand{\sqrg}{\sqrt{\g\hspace{0.5pt}}\hspace{0.5pt}\vphantom{|}}%{\sqrt{{\!\,\!{\shortmid}{\g}{\shortmid}\!\,\!}}}%{} %{{\sqrt{\g}}}

\newcommand{\lapse}{\!\tskip\scriptscriptstyle{\perp}\!}

\newcommand{\INs}[1]{{#1}}  %# shift
\newcommand{\INl}[1]{{#1}^{\lapse}}  %# lapse

\newcommand{\iNs}[1]{{#1}}  %# shift
\newcommand{\iNl}[1]{{#1}_{\lapse}}   %# lapse

\newcommand{\Ns}{\INs{N}}
\newcommand{\Nl}{\INl{N}}
\newcommand{\lmrNl}{\iNl{\lambda}}

\newcommand{\zk}{{k}}   %k
\newcommand{\zl}{{l}}   %l
\newcommand{\zm}{{m}}   %m
\newcommand{\zn}{{n}}   %n

\newcommand{\Zm}{{\mu}}      %m
\newcommand{\Zn}{{\nu}}      %n

\newcommand{\var}{\operatorname{\delta}\hspace{-1pt}} %{\hspace{0.8pt}\delta} % variation symbol

\newcommand{\TDer}[2]{\frac{{d}\tskip{#1}} {{d}\tskip{#2}}}  % total derivative

\newcommand{\VDer}[3][]{ \frac{{\var^{#1}}  {#2}} {{\var^{}} {#3}} }  % variational derivative

\newcommand{\hmg}[1]{\overline{#1}} %{\bar{#1}}
\newcommand{\inh}[1]{\widetilde{#1}} %{\check{#1}} {\tilde}

\newcommand{\geps}{\ColorGauge{\varepsilon}} %additional grouping makes accent (bar, dot) position incorrect (lost italics adjustment)
\newcommand{\gxi}{\ColorGauge{\xi}} %additional grouping makes accent (bar, dot) position incorrect (lost italics adjustment)
\newcommand{\geta}{\ColorGauge{\eta}}
\newcommand{\gzeta}{\ColorGauge{\zeta}}

\DeclareMathOperator{\eomeq}{{\,\accentset{{\color{GhostWhite}\sim}}{=}\,}}
\DeclareMathOperator{\defeq}{{\,\accentset{{\color{GhostWhite}---}}{=}\,}}

\newcommand{\FF}{F}%{\scalebox{1.1}{$v$}}
\newcommand{\WW}{W}%{\scalebox{1.1}{$w$}}

\newcommand{\wwc}{\mathrm{\hspace{0.5pt}w}}

\newcommand{\OOmega}{\Omega}
\newcommand{\TTheta}{\Theta}

\newcommand{\pg}{\pi}%{p}             % 3-metric momenta
\newcommand{\trpg}{\pg} %{{\color{Brown}\pg}}%        % 3-metric momenta trace

\newcommand{\Hs}{\iNs{H}}                        % Momentum constraint
\newcommand{\Hl}{\iNl{H}} %{H_{^{_{\!\perp}}\!}}      % Hamiltonian constraint

\newcommand{\lmCT}{\mu} %% vector Lagrange multipliers at \CT_{,\zm} %% 2024 instead of deprecated \lmv

\newcommand{\CPI}{P_{{\scriptscriptstyle{I}}}}

\newcommand{\tf}{\rep{\tau}}
\newcommand{\ptf}{\rep{\pi\hspace{-0.6pt}}}  % canonically conjugated momenta for time-field
\newcommand{\replm}{\rep{\lambda}}  % parameterization Lagrange multiplier

\newcommand{\GBulk}{g} %{\textsl{g}} {\textbf{g}}  %# -- bulk (spacetime) metric

\newcommand{\sR}{{\color{sColor} \HIDE{^{^{\sdim}\!\!}}R}}  % spatial curvature
\newcommand{\stR}{{^{^{\Ddim}\!\!}R}\vphantom{|}}  % spacetime curvature

\newcommand{\iGUMG}{} %{{\scriptscriptstyle G\!\hspace{0.5pt}U\!M\!G}}
\newcommand{\iwwc}{{\scriptscriptstyle{(\!\wwc\!)}}}

\newcommand{\io}{{\scriptscriptstyle{0}}}

\newcommand{\ibo}{{\scriptscriptstyle{\color{igray}\mathbf{0}}}}  % bold index 0

\newcommand{\gzetat}{\gzeta_{{\color{mygray}(\hspace{-1pt}t\hspace{-1pt})}}}

\DeclareRobustCommand{\op}[1]{\hat{#1}}

\newcommand{\argsqrg}{{\scalebox{0.8}{\mbox{\!\ensuremath{\sqrg}}}}}

\newcommand{\rep}[1]{\ColorRepAux{#1}} %%{\breve{#1}}

\newcommand{\Xaux}{\rep{X}}

\newcommand{\taux}{\rep{t}} %$\tau$ time variable in parameterized action

\if{
}\fi

\newcommand{\replmCT}{\rep{\lmCT}}%{\rep{\lmv}}
\newcommand{\lmptf}{\rep{\nu}}%{\rep{\lmv}}

 \newcommand{\HTf}{\rep{\mathcal{V}}} %{\rep{\mathcal{X}}}

 \newcommand{\CCf}{\rep{\varLambda}}  %{\varLambda\hspace{0.5pt}}
 \newcommand{\CCof}{\rep{\varLambda}_\ibo}  %{\varLambda\hspace{0.5pt}}
 \newcommand{\cco}{\mathrm{c}}

\newcommand{\Gaux}{\rep{g}}%{\rep{\GG}}

\newcommand{\repNs}{\rep{\Ns}}%{\rep{\Ns}}
\newcommand{\repNl}{\rep{\Nl}}%{\rep{\Nl}}
\newcommand{\ssep}{\hspace{0.4mm}{\textrm{,}}\hspace{0.4mm}}
\def\tauxdot{\dot} %{\mathring}

\newcommand{\wGUMG}{$\wwc$-\GUMG}
\newcommand{\Fw}{F_{\!\iwwc}}

\newcommand{\inhId}{\mathop{\inh{\mathbb{I}}}}
\newcommand{\hmgId}{\mathop{\hmg{\mathbb{I}}}}

\newcommand{\ofa}{a}

\newcommand{\sDvg}[2][]{\partial_{#1}#2^{#1}}

\newcommand{\TSEt}{T}
\newcommand{\pSEt}{p}
\newcommand{\eSEt}{\varrho} %{\varepsilon}

\newcommand{\repTSEt}{\rep{T}}
\newcommand{\reppSEt}{\rep{p}}
\newcommand{\repeSEt}{\rep{\varrho}} %{\varepsilon}

\newcommand{\teq}{{\,=\,}} %% text mode =
\newcommand{\tequiv}{{\,\equiv\,}} %% text mode \equiv

\def\tauxdot{\dot} %{\mathring}

\begin{document}
% Use the \preprint command to place your local institutional report
% number in the upper righthand corner of the title page in preprint mode.
% Multiple \preprint commands are allowed.
% Use the 'preprintnumbers' class option to override journal defaults
% to display numbers if necessary
%\preprint{}

%Title of paper
\title{Henneaux-Teitelboim Form of the Generalized Unimodular Gravity Action
 %\\ \phantom{.}
 \vspace{0.5mm} }
%\vspace{1mm}

% repeat the \author .. \affiliation  etc. as needed
% \email, \thanks, \homepage, \altaffiliation all apply to the current
% author. Explanatory text should go in the []'s, actual e-mail
% address or url should go in the {}'s for \email and \homepage.
% Please use the appropriate macro foreach each type of information

% \affiliation command applies to all authors since the last
% \affiliation command. The \affiliation command should follow the
% other information
% \affiliation can be followed by \email, \homepage, \thanks as well.
  \author{Dmitry~Nesterov} %$^{\star}$  %D.V.~Nesterov
\email[]{nesterov@lpi.ru}
%\homepage[]{Your web page}
%\thanks{}
%\altaffiliation{}
  \affiliation{ \HIDE{I.E.Tamm }Theory Department, P.N. Lebedev Physical Institute of the Russian Academy of Sciences,  Leninsky Prospekt 53, Moscow, 119991, Russia} % $^{\star}$

  \author{Ksenia~Lyamkina}  %$^{\dag}$ %K.~Lyamkina
%\email[]{}
%\homepage[]{Your web page}
%\thanks{}
%\altaffiliation{}
  \affiliation{Moscow Institute of Physics and Technology, Institutskiy per.~9, Dolgoprudny,~Moscow Region, 141700, Russia} % $^{\dag}$

%Collaboration name if desired (requires use of superscriptaddress
%option in \documentclass). \noaffiliation is required (may also be
%used with the \author command).
%\collaboration can be followed by \email, \homepage, \thanks as well.
%\collaboration{}
%\noaffiliation

%\date{\today}

% insert abstract here
\begin{abstract}
\vspace{1.3mm}

 %% We propose an alternative formulation of generalized unimodular gravity ({\GUMG}), extending the Henneaux-Teitelboim approach for unimodular gravity ({\UMG}).  The central feature of this formulation is the consistent incorporation of time parameterization into the theory, which extends its gauge structure and uncovers an underlying spatial nonlocality hidden in the dynamics of the original formulation. We briefly discuss the resulting dynamics, including the effect of spatial nonlocality, and outline the extended gauge structure, which is not yet\HIDE{ fully} diffeomorphism-invariant. Furthermore, we identify a subfamily of {\GUMG} models whose alternative description remains manifestly local.

 %We propose an alternative description of generalized unimodular gravity ({\GUMG}), extending the Henneaux-Teitelboim approach to unimodular gravity ({\UMG}).  The central feature of this formulation is the consistent incorporation of time reparameterization, which enhances the gauge structure and reveals a spatial nonlocality hidden in the dynamics of the original formulation. We examine the resulting dynamics, emphasizing the effects of spatial nonlocality, and outline the constraint structure. In particular, we show that the gauge symmetry in the gravitational sector is extended by a functionally incomplete symmetry, as occurs in the unimodular gravity. However, in contrast to the latter, the resulting action is not fully diffeomorphism-invariant. Furthermore, we identify a subset of {\GUMG} models for which the alternative formulation preserves manifest locality.

 We propose an alternative description of generalized unimodular gravity ({\GUMG}), extending the Henneaux-Teitelboim approach to unimodular gravity ({\UMG}).  The central feature of this formulation is the consistent incorporation of time reparameterization, which enhances the gauge structure and reveals a spatial nonlocality hidden in the dynamics of the original formulation. We examine the resulting dynamics, emphasizing the effects of spatial nonlocality, and outline the constraint structure. We show that the gauge symmetry in the gravitational sector is extended by a functionally incomplete symmetry, as occurs in the unimodular gravity. However, in contrast to the latter, the resulting action is not fully diffeomorphism-invariant. %Furthermore, we identify a subset of {\GUMG} models for which the alternative formulation preserves manifest locality.

\vspace{1.3mm}
\phantom{.}
\end{abstract}

% insert suggested keywords - APS authors don't need to do this
%\keywords{}

%\maketitle must follow title, authors, abstract, and keywords
\maketitle

% body of paper here - Use proper section commands
% References should be done using the \cite, \ref, and \label commands
%\section{...}
% Put \label in argument of \section for cross-referencing
%\section{\label{}}

%%-------------------------=%        ******         %=-------------------------%%
%%-------------------------=%        ******         %=-------------------------%%
%%-------------------------=%        ******         %=-------------------------%%

%\vspace{1mm}

\section{Generalized unimodular and unimodular gravity theories}
\vspace{1mm}

Modifications of general relativity are an active area of research, driven by\HIDE{ attempts to address} open problems in cosmology, quantum gravity, puzzles of dark energy and dark matter.
Among these modifications, it is worth noting
models that deform Einstein's general relativity (\GR)
by imposing additional restrictions on the configuration
space.
A prominent example of restricted gravity theories is unimodular gravity (\UMG) with the local restriction $\!\sqrt{|\GBulk|} \teq 1 $ on the spacetime metric determinant \cite{Einstein1919UMG,Weinberg:1988cp,Unruh:1988in,Ng:1990xz,Bousso:2007gp,Bufalo:2015wda}.
Although {\UMG} preserves the local physical dynamics of general relativity, it introduces subtle changes by constraining the spacetime volume and leaving the cosmological constant undetermined. A {virtue}\HIDE{ notable advantage} of this theory is the alternative covariant formulation proposed by Henneaux and Teitelboim \cite{Henneaux:1989zc}.

Generalized unimodular gravity (\GUMG) models, %defined by the action
introduced by Barvinsky and Kamenshchik\HIDE{ in} \cite{Barvinsky:2017pmm},
  \bea{Action_GUMG_L} % (\ref{Action_GUMG_L})
    \!
    S^{\iGUMG}[\GBulk,\lmrNl]
    = \! \int \hspace{-1pt}\!d \tx \hspace{1pt} \dsx \,
   \LiB\!
     \sqrt{|\GBulk|}\,\stR(\GBulk)
     -
     \lmrNl \big(\Nl {-\,} \FF(\argsqrg) \big)
   \RiB \hspace{-1pt}
    ,\;\;\;\;
  \eea
 %%were initially motivated by search for new\HIDE{ viable} descriptions of dark energy,
\HIDE{structurally} are the simplest generalizations of unimodular gravity.
In the action (\ref{Action_GUMG_L}), $\GBulk_{\Zm\Zn}$ is the unrestricted $\Ddim$-dimensional spacetime metric, decomposed into Arnowitt–Deser–Misner (ADM) components \cite{Arnowitt:1959ah,Misner:1973prb} as
  $\vphantom{\big|^i_{l_l}}
  %\beq{GBulk_metrics_ADM}
   \GBulk_{\Zm\Zn}d{X}^{\Zm}d{X}^{\Zn}
    {\,=\,\,}  \g_{\zm\zn} (d\sx^\zm {+} {\Ns}^\zm d\tx)(d\sx^\zn {+} {\Ns}^\zn d\tx) - {\Nl}^2 d\tx d\tx
    ,%\vphantom{\big|}
  %\eeq
  $
where $\Nl$ is the lapse function, $\Ns^\zm$ are shift functions, and $\g_{\zm\zn}$ is the induced metric on $\,\tx \teq const\,$ hypersurfaces. The first term in (\ref{Action_GUMG_L}) is the {\GR} Einstein-Hilbert action and the second term via the Lagrange multiplier mechanism enforces the {\GUMG} restriction condition,
  \beq{GUMG_restriction}
    \Nl \eomeq \FF(\argsqrg)
    \,,
  \eeq
which expresses the lapse function\HIDE{ $\Nl$} in terms of some positive monotonic function of the square root of the induced-metric determinant \,$\sqrg \defeq \sqrt{\det{\g_{\zm\zn}}}$. The choice of $\FF\HIDE{(\argsqrg)}$ defines a specific model in the {\GUMG} family.

\vphantom{$\big|^I$}The observation, which\HIDE{ initially} drew attention to {\GUMG}\HIDE{ models}, was that the restriction term in (\ref{Action_GUMG_L}) generates an energy-momentum tensor of the cosmological {perfect fluid},
  $ % \bea{GUMGrStressTensor}
    \, \TSEt^{\Zm\Zn}
     {\,=\,\,} \eSEt \,u^\Zm u^\Zn + \pSEt \,(\GBulk^{\Zm\Zn} {+\,} u^\Zm u^\Zn)
     \,,\,
  $ %  \eea
on the right\HIDE{-hand} side of the Einstein equations
($u^\Zm$ --- timelike unit vector field, normal to $\tx=const$ hypersurfaces). Expressions for energy density, $\eSEt \,\teq \frac{\lmrNl\vphantom{|}}{2\sqrg}$, and pressure, $\pSEt\, \teq \frac{\lmrNl\vphantom{|}}{2\sqrg}\frac{\FF\WW}{\Nl}$,
on the restriction surface\HIDE{ $\Nl \!= \FF(\argsqrg)$,} (\ref{GUMG_restriction}) imply the equation of state\HIDE{ of the cosmological perfect fluid}
  \bea{PerfectFluid_Eq_of_state}
    \pSEt \eomeq  \WW \eSEt
    \,,
  \eea
where barotropic equation-of-state parameter,
  \bea{def_WW_GUMG}
    \WW(\argsqrg) \defeq \TDer{\,\ln \FF(\argsqrg)}{\,\ln{\sqrg}}
    \,,
  \eea
is the derived characteristic function which depends on the induced volume density on spatial sections
 \footnote{To avoid pathologies in the\HIDE{ canonical} constraint structure, we assume $\WW(\argsqrg)$ to be a sign-definite function\HIDE{ in the domain $\sqrg>0$}, which implies monotonicity of $\FF(\argsqrg)$. However, the\HIDE{ interesting }{\HIDE{cold }dust} case, $\WW\equiv0$,\HIDE{ also} fits the description.}.

\vphantom{$\big|^I$}Besides initial motivation for a new dark energy mechanism, the dynamical and constraint analysis of {\GUMG} \cite{Barvinsky:2019agh,Barvinsky:2019qzx} reveals additional features, such as the possibility of inflationary scenario\HIDE{ which is} consistent with observations\HIDE{ \cite{Barvinsky:2019qzx}} for a certain subclass of $\FF$.
The dynamical nature of $\WW$ is also an attractive feature in cosmological phenomenology.

\vphantom{$\big|^I$}Unimodular gravity emerges as a special case of the {\GUMG} family, defined by
  \beq{UMG_part_case}
   \FF(\argsqrg)=\sqrg^{-1},
   \quad \WW(\argsqrg) \equiv - 1
   \,.\quad \;\; \text{[\UMG]} \quad
  \eeq
However, this is an exceptional case, whose dynamical content is almost general-relativistic, whereas {\GUMG} describes gravitational dynamics in the presence of additional physical degree of freedom associated with the cosmological perfect fluid \cite{Barvinsky:2017pmm,Barvinsky:2019agh}.
Also, unlike unimodular case, {\GUMG} models imply a preferred time direction, reflecting the space-time anisotropy of the prefect fluid.

%%-------------------------=%        ******         %=-------------------------%%
%%-------------------------=%        ******         %=-------------------------%%
%%-------------------------=%        ******         %=-------------------------%%

%\vspace{17.8mm}

\section{{\GUMG} analog of the Henneaux-Teitelboim action for {\UMG}}

Henneaux and Teitelboim\HIDE{ in 1989} suggested an alternative formulation of unimodular gravity via the so-called covariant action \cite{Henneaux:1989zc},
providing unrestricted diffeomorphism invariance while remaining physically equivalent to the standard {\UMG} setup\HIDE{ as a restricted {\GR}}.
%% representation in the form of restricted {\GR}.

We claim that analog of the Henneaux-Teitelboim representation for {\GUMG} theories
is
 \bea{ActionHTlike_GUMG} %(\ref{ActionHTlike_GUMG})
    S^{\iGUMG}_{\ialt}[\Gaux,\CCf,\HTf]
    &=&  \int \!d\taux \hspace{1pt} \dsx \, \sqrt{|\Gaux|} \,\big( \stR(\Gaux) - \CCf \big)
    \nonumber \\
    && +  \!\int \!d\taux \hspace{1pt} \dsx \, \partial_\Zm \HTf^\Zm  \op{E}  \FF  \sqrg \CCf
    \,, \quad
 \eea
where $\stR(\Gaux)$ --- scalar curvature for unrestricted spacetime metric $\Gaux_{\Zm\Zn}$,
$\sqrg$ is the covariant induced volume density on $\taux \teq const$ hypersurfaces,
$\CCf$ is the so-called {cosmological-constant field},
$\HTf^\Zm $ is the auxiliary Henneaux-Teitelboim field, and $\op{E}$ is the spatially nonlocal, nondegenerate operator defined by the kernel
  \bea{opE_kernel}
    E(\taux|\sx;\sx')
    \,=\, \delta(\sx;\sx')
    - \frac{\WW^{-1} (\taux,\sx) \, \WW (\taux,\sx')}{\sVol\HIDE{(\taux)}}
    + \frac1{\sVol\HIDE{(\taux)}}
    \,,\;\;
  \eea
where $\WW(\taux,\sx) \defeq \WW(\sqrg {\scalebox{0.8}{\mbox{\ensuremath{(\taux,\sx)}}}})$ and $\sVol\HIDE{(\taux)} \defeq \sint 1$ is the background spatial volume. Integrals without\HIDE{ specifying} explicit measure denote \NBT{spatial} integration
  \beq{}
    \sint f (\taux) \defeq \textstyle{\int\limits_{{\scriptstyle\!\taux=const\!}}} \!\!\!\! d\sx\, f (\taux,\sx) \,.
\vspace{-3mm}
  \eeq
In the unimodular case (\ref{UMG_part_case}) the whole factor $ \op{E}  \FF  \sqrg$ becomes unity and the {\UMG}\HIDE{ Henneaux-Teitelboim} covariant action \cite{Henneaux:1989zc} is restored.

Another convenient form of the action, %% for analyzing dynamics and gauge structure,
 \bea{ActionHTlike_GUMG_ccf0}
    S^{\iGUMG}_{\ialt_0}[\Gaux,\CCof,\HTf]
    &=& \! \int \!d\taux \hspace{1pt} \dsx \, \sqrt{|\Gaux|} \Big( \stR(\Gaux) - \sqrg^{-1}\!\FF^{-1}\!\op{E}^{-1}\!\CCof \Big)
    \nonumber\\
    && +  \!\int \!d\taux \hspace{1pt} \dsx \, \partial_\Zm \HTf^\Zm  \CCof
    \,, \quad
   % \nonumber
  \eea
is obtained from (\ref{ActionHTlike_GUMG}) via nondegenerate redefinition
  $ % \beq{ccf_0_GUMG}
   \CCof
   {\defeq} \op{E} \FF \sqrg \CCf
   \HIDE{= \ptf_\io}
   \,.
  $ %\eeq
%% Field $\CCof$ is the gauge-invariant representation of the cosmological-constant field,\HIDE{ which is an unfixedspacetime} constant on shell.
The inverse operator $\op{E}^{-1}$ has the kernel
  \bea{invE_kernel}
    E^{-1}(\taux|\sx;\sx')
      &=& \delta(\sx;\sx')
      -  \frac{\WW^{-1}(\taux,\sx)}{\sint {\WW^{-1}}(\taux)}
      - \frac{\WW(\taux,\sx')}{\sint{\WW}(\taux)}
     \nonumber\\
     &&\qquad
      + \,2\,
        \frac{\WW^{-1}(\taux,\sx)\, \sVol\HIDE{(\taux)}\, \WW(\taux,\sx')}{\sint{\WW^{-1}(\taux)}\;  \sint{\WW}(\taux)}
    \,.
  \eea
The convenience follows from the gauge invariance of $\CCof$ field, which is a spacetime constant on shell.

The origin and consequences of the delocalization will be discussed later. Here we highlight a prominent {\GUMG} subfamily, termed {\wGUMG} and defined by
  \beq{AMG_part_case}
   %\text{\wGUMG: }\quad
   \!\!\FF(\argsqrg)=\sqrg^{\wwc},
   \quad
   \!\!\WW(\argsqrg)\equiv \wwc = const
   \,,
   \quad
   \!\!\!
   \text{[\wGUMG]}
  \eeq
for which spatial delocalization disappears because $\op{E}$ and $\op{E}^{-1}$ become identity operators with kernel $\delta(\sx,\sx')$. \footnote{The\HIDE{ constructive} derivation admits a freedom in the form of the delocalization: for any sign-definite\HIDE{ function} $\ofa(\g_{\zm\zn})$, operators
 $ %\beq{opE_kernel}
  E_\ofa(\taux|\sx;\sx') {\,\,=\,\,} \ofa^{-1}\! (\taux,\sx) \delta(\sx;\sx') \WW (\taux,\sx')
  - \ofa^{-1}\! (\taux,\sx) \tfrac1{\sVol\HIDE{(\taux)}}  \WW (\taux,\sx')
  + \tfrac1{\sVol\HIDE{(\taux)}} \HIDE{\sVol^{-1}}
 % \,,
 $ % \eeq
 %%with any sign-definite\HIDE{ function} $\ofa(\g_{\zm\zn})$
 describe classically equivalent models. The choice (\ref{opE_kernel}) is preferable
  %%not only due to simplifications, but also in view of
 mainly because of the identity limit for {\wGUMG}.
 %%\HIDE{ and better asymptotic properties in noncompact space.}
}

%%-------------------------=%        ******         %=-------------------------%%
%%-------------------------=%        ******         %=-------------------------%%
%%-------------------------=%        ******         %=-------------------------%%

\section{Time parameterization of  {\GUMG} canonical action}

In this section, we derive the alternative {\GUMG} action, which simultaneously establishes the\HIDE{ physical} equivalence of the representations. The key step\HIDE{ leading to Henneaux-Teitelboim form of the action} is the consistent time parametrization of the canonical extended action of the theory, which deserves special discussion.

We start with the extended canonical action, previously analyzed for general {\GUMG} in \cite{Barvinsky:2019agh} using the Dirac scheme. Here, we briefly revisit this through a concise hamiltonization. First, we reduce (\ref{Action_GUMG_L}) with respect to the\HIDE{ pair of auxiliary variables} $\lmrNl$, ${\Nl}$, then introduce conjugate momenta $\pg^{\zm\zn}$ only for the spatial metric $\g_{\zm\zn}$\HIDE{ via a Legendre transform}. This yields a canonical theory on the phase space $(\g_{\zm\zn} \ssep \pg^{\zm\zn})$ with a Hamiltonian density $\FF(\argsqrg) \Hl(\g,\pg)$ and primary constraints $\Hs_\zn(\g,\pg) \teq 0$, corresponding to spatial diffeomorphisms. The structures $\Hl\HIDE{(\g,\pg)}$ and $\Hs_\zn\HIDE{(\g,\pg)}$ are identical to the Hamiltonian and momentum constraints in canonical Einstein gravity \cite{Misner:1973prb},
 \bea{}
    {\Hl}(\g,\pg) &=& \tfrac{1}{\sqrg} \left(\pg^{\zm\zn}\pg_{\zm\zn} \!- \tfrac{1}{\Ddim{-}2}\trpg^2\right) - \sqrg \sR(\g)
    \,,
    \label{BGUMGHamiltonianStructure}
    \\
    {\Hs}_\zn(\g,\pg)  &=& -2\g_{\zn\zm} {\pg^{\zm\zk}}_{;\zk} \,,
    \label{BGUMGMomentaConstraints}
\eea
where $\pg_{\zm\zn} {\defeq} \g_{\zm\zk}\pg^{\zk\zl}\g_{\zl\zn}$, \, $\trpg {\defeq} \g_{\zm\zn}\pg^{\zm\zn}$,\, $\sR(\g)$ is the scalar curvature of $\tx \teq const$ hypersurfaces, and semicolons denote spatial covariant derivatives. It is important to note that in {\GUMG} ${\Hl}$ (\ref{BGUMGHamiltonianStructure}) is not a constraint, but a factor in the nontrivial Hamiltonian density $\FF(\argsqrg) \Hl(\g,\pg)$.

Dynamics do not preserve\HIDE{ the spatial diffeomorphisms primary} primary constraints $\Hs_\zn \!\teq 0$ (for $\WW{\,\not=\,}0$, \footnote{The {\HIDE{cold }dust} case, $\WW \tequiv 0$ ($\FF \teq const$),  is the special {\GUMG} case in which dynamics preserves all\HIDE{ spatial diffeomorphism} constraints $\Hs_\zn\!\teq 0$, which are all first-class. However, the constraint $\big(b(\argsqrg)\Hl\big)_{,\zn}$ can be added by hand as a partial gauge fixing. Thus, the following derivation and the final representation are also valid for $\WW \tequiv 0$ case upon substituting $\WW(\argsqrg)\to b(\argsqrg)$ with almost any $|b(\argsqrg)|>0$.
 })
and the Dirac consistency procedure reveals\HIDE{ functionally incomplete the} secondary constraints
  \bea{secondary_GUMG}%{tertiary_GUMGr}
   %% \big( \WW\FF\Hl \big)_{,\zm}(\g,\pg) = 0
   \big( \WW(\argsqrg) \FF(\argsqrg) \Hl(\g,\pg) \big)_{,\zm} = 0
   \,,
  \eea
which have a form of spatial gradient ($f_{,\zm}\HIDE{(\tx,\sx)}{\defeq\,} \partial_\zm f\HIDE{(\tx,\sx)}$). This effectively constrains the average-free component
  %% and in view of the assumed compactness of $t{\,=\,}const$ hypersurfaces are equivalent to the average-free scalar constraint
$ \inh{\WW\FF\Hl}  \tequiv\,  \WW\FF\Hl \!- \hmg{ \WW\FF\Hl } \,\teq 0$ on $\,\tx \teq const\,$ hypersurfaces.
We denote \NBT{average} ({\HIDE{spatially }homogeneous}) and \NBT{average-free} ({inhomogeneous}) components of local functions with overlines and tildes respectively,
 \footnote{Various arguments rely on decomposing quantities on average and average-free parts\HIDE{ via complementary projectors $\hmgId$ and $\inhId$} and on the equivalence of average-free quantities with spatial vector divergencies. Though these are straightforward for compact spaces\HIDE{ without boundaries}, the construction extends for spacetimes with noncompact spatial sections \cite{AltActionFull:2024}.
 \if{
   %% For example, by separating background nonintegrable contributions from certain fields and performing the analogous projecting for the sequence of measures, which in the limit correspond to integrating over the whole noncompact spatial section.
   %%In fact, only the existence of such decomposition and the equivalence of average-free quantities with spatial vector divergencies are essential \cite{AltActionFull:2024}.
   %% These issues are discussed and justified in \cite{AltActionFull:2024}.
   %% In the present discussion we only exploit the existence of such projections and the possibility to represent average-free functions as spatial vector divergencies (and vice versa).
   %% And the possibility not to control the total spatial derivatives when integrating by parts.
 }\fi
}
 \bea{}
  \hmg{f}(\tx) \defeq \sint f(\tx,\sx) / \sint 1
  \,,
  \quad \;\;
  \inh{\,f\,\,}\!(\tx,\sx) \defeq f(\tx,\sx) - \hmg{f}(\tx)
  %\quad  \sVol\HIDE{(t)} \equiv
  \,.
  \;\;\;
 \eea

The constraints $\Hs_\zn \HIDE{(\g,\pg)} {=\,} 0$ and $(\WW\FF\Hl)_{,\zm}\HIDE{(\g,\pg)} {=\,} 0$ form a complete set\HIDE{ (\ref{Barvinsky:2019agh})}
\footnote{We do not provide a separate discussion\HIDE{ in the main text} of the\HIDE{ peculiarities associated with the} so-called GR branch\HIDE{ (``gauge'', ``parental'')} of {\GUMG} \cite{Barvinsky:2019agh}, which is the $\Hl=0$ submanifold\HIDE{ of the physical space,} characterized by a rank discontinuity of the matrix of Poisson brackets of constraints. Changes affect the constraint classification: on the GR branch all constraints become first-class and thus there are no dynamical constraints on any of the Lagrange multipliers. Importantly, the GR branch is\HIDE{ preserved and} correctly described by the\HIDE{ presented actions of the} new parameterized {\GUMG} description.
},
so the {\GUMG} concise extended\HIDE{ canonical} action reads
%\vspace{-1mm}
 \bea{extAction_GUMG} % (\ref{extAction_GUMG})
    S_{{\scriptscriptstyle{\!E}}}^{}[\g,\pg,\Ns,\lmCT]
    &=& \int\! d\tx \hspace{1pt} \dsx \, \LiB
     \dot{\g}_{\zm\zn} \pg^{\zm\zn}
     - \FF \Hl
     \nonumber\\
     && \qquad
     - \Ns^\zn \Hs_\zn
     - \lmCT^\zm (\WW\FF\Hl)_{,\zm}
    \RiB
    .\;\;
  \eea

\vspace{1mm}

Time parameterization of the field-theoretic canonical action in a mechanical-like form by means of the local fields, which was applied in \cite{Henneaux:1989zc}
for {\UMG}, remains valid for {\wGUMG} subfamily, characterized by constant $\WW \tequiv \wwc \teq const\,$ and $\FF \to \Fw \teq \sqrg^{\wwc}$,\, \HIDE{(\ref{AMG_part_case}),} for which the\HIDE{ equivalent} parameterized action
  %%\footnote{The particular interpretation of time parameterization procedure --- either a new auxiliary time is introduced and initial time coordinate is encoded in the field $\tf^\io$, or just new auxiliary fields are introduced on initial spacetime --- is not important, because both these points of view imply physical equivalence of the parameterized theory and initial canonical model. This is the only fact we need here.}
can be written as
  \bea{paramAction_naive_AMG}
    S^{\iwwc}_{par} [\g,\pg,\tf^\io\!,\ptf_\io,\replm^\io\!,\repNs,\replmCT]
    &=&\! \int\! d\tx \hspace{1pt} \dsx \, \LiB
     \tauxdot{\g}_{\zm\zn}\pg^{\zm\zn}\!
     + \tauxdot{\tf}^\io\ptf_\io
     \nonumber\\
     &&
     \hspace{-90pt}
     - \replm^\io ( \ptf_\io {+} {\Fw \Hl} )
     - \repNs^\zn \Hs_\zn
     - \replmCT^\zm (\wwc \Fw\Hl)_{,\zm}
     %%- \replmCT^\zm {\CTw}_{,\zm}
     \RiB
     . %,
   \qquad% \nonumber\\
  \eea

% - - - - -
%\vspace{2mm}

To ensure the equivalence of the original and parameterized representations of general field theory, two conditions must be satisfied.
 First, the new parameterized Hamiltonian constraint must introduce a new \NBT{first-class} constraint into the extended set. This new constraint is not necessarily the parameterized Hamiltonian itself, it may involve a linear combination with the original constraints.
 Second, there must exist an \NBT{accessible} gauge fixing (e.g. ${\tf}^\io{-\,}\taux \teq 0$) for the new first-class constraint, which recovers the original\HIDE{ extended} action (\ref{extAction_GUMG}) upon reduction.

These requirements are satisfied in {\wGUMG} and {\UMG} for the local parameterized action (\ref{paramAction_naive_AMG}).
Notably, in {\wGUMG} ($\wwc{\,\neq\,}{-}1$) the new first-class constraint introduced through parameterization is
 $ \sint \geps \CPI
   \defeq \sint \geps (\ptf_\io{+\,}\hmg{\Fw\Hl})
   \,\teq \sint (\geps \ptf_\io{+\,}\hmg{\geps}\,\Fw\Hl)
   ,
 $
which involves only the average component of the Hamiltonian density.
(By $\geps$, $\gzeta$, $\geta$, $\gxi$, etc. we denote local gauge parameters.)
This mean that the time reparametrization gauge symmetry introduced in (\ref{paramAction_naive_AMG}) by local parameterization, in the sector of metric fields manifests as a functionally incomplete symmetry with a homogeneous\HIDE{ gauge} parameter $\hmg{\geps}(\taux)$. Whereas in the sector of the auxiliary fields this symmetry remains local.
The constraint $ \sint \geps \CPI$ is a linear combination of the local parameterization constraint $\sint \geps (\ptf_\io{+}{\Fw\Hl})$ and the secondary constraint $\sint(\partial_\zn \geta^\zn) (\wwc\Fw\Hl)$ with $(\partial_\zn \geta^\zn) \teq {-}\inh{\geps}\,\wwc^{-1}$. Both of which are second class.
 %%\HIDE{ because of nontrivial Poisson bracket with the longitudinal component of $\Hs_\zn$}
This suggests that the local\HIDE{ mechanical-like} parameterization (\ref{paramAction_naive_AMG}) is valid because the
 %%\HIDE{ due to constancy of $\WW\equiv\wwc$,}
average-free part of the Hamiltonian density was constrained (\ref{secondary_GUMG})\HIDE{ in the action (\ref{extAction_GUMG})} prior to parameterization
\if{
 \footnote{
 In \cite{Henneaux:1989zc} parameterization was aimed at completing the $\Hl$ structure, which lacks a mode in the secondary constraint. When the new constraint is introduced the,\HIDE{ canonical} {\GR}-like constraint structures can be extracted in an appropriate constraint basis. However, the desired completion\HIDE{ of the Hamiltonian density} can be achieved by only the \NBT{homogeneous} parameterization by means of the spatially homogeneous auxiliary fields $\hmg{\tf}^\io(\taux), \hmg{\ptf}_\io(\taux), \hmg{\replm}^\io(\taux)$ \cite{Henneaux:1989zc}, which introduces $ \HIDE{\sVol} \big( \partial_{{\taux}}{\hmg{\tf}^\io} {\!\cdot}\hmg{\ptf}_\io - \hmg{\replm}^\io ( \hmg{\ptf}_\io {+} \hmg{\Fw \Hl} )\big)$ into the action integrand.
 The local action (\ref{paramAction_naive_AMG}) can be obtained from homogeneously parameterized action by adding pure gauge combination of complementary average-free fields\HIDE{ (and rearranging constraints)}.
 %%The true advantage of the local parameterization (\ref{paramAction_naive_AMG}) is the local nature of auxiliary fields.
}.
}\fi
 %
%{\color{Gray}
\footnote{
 The constraint basis of the parameterized {\wGUMG} can be rearranged to disentangle first-class subspace spanned by $\HIDE{\int \geps} \CPI$ and the transverse part of $\HIDE{\int \gzetat^\zn} \Hs_\zn$,\HIDE{ ($\partial_\zk \gzetat^\zk{=\,}0$),} which generate the canonical gauge algebra of the model. Meanwhile, the average-free part of ${\Fw\Hl}$ (encoded in the secondary constraint\HIDE{ $(\wwc\Fw\Hl)_{,\zn}=0)$}) and the longitudinal part of $\Hs_\zn$ form a pair of second-class constraints.
 In the exceptional {\UMG} case, $\wwc{\,=\,}{-}1$\HIDE{ (\ref{UMG_part_case})}, this pair\HIDE{ of constraints} becomes first-class, restoring full spacetime diffeomorphism invariance\HIDE{ in the metric sector} in accordance with \cite{Henneaux:1989zc}.
}.

 %% short ver:
 %% In general field theory, as well as for {\GUMG} with nonconstant $\WW(\argsqrg)$, the mechanical-like parametrization analogous to (\ref{paramAction_naive_AMG}) \NBT{does not} produce an equivalent theory\HIDE{ and thus parameterization procedure should be modified}. \HIDE{This occurs because }The local constraint $\ptf_\io{+}\FF\Hl$ fails to introduce a new first-class constraint: the Poisson bracket matrix for a set $\big(\ptf_\io{+}\FF\Hl \ssep \Hs_\zn \ssep (\WW\FF\Hl)_{,\zm}\big)$ has greater rank than for the initial set $\big(\Hs_\zn \ssep (\WW\FF\Hl)_{,\zm}\big)$. To maintain the rank and satisfy the correspondence condition, one may extend all constraints with $O(\tf^\io_{,\zn})$ terms. While a solution can be formally constructed, it is generally impractical\HIDE{ for most of field theories,} as it results in an infinite series in powers of $\tf^\io$  and its spatial derivatives.

 %% long ver:
 In the general field theory, particularly for {\GUMG} with nonconstant $\WW(\argsqrg)$, a mechanical-like parametrization similar to (\ref{paramAction_naive_AMG}) \NBT{does not} yield an equivalent theory\HIDE{ and thus parameterization procedure should be modified}. %%This is common situation in field theories.
 \HIDE{This occurs because }  The local constraint $\ptf_\io{+}\FF\Hl$ fails to introduce a new first-class constraint. From the Poisson brackets
 \newcommand{\PB}[2]{{\mathchoice{\big\{{#1},{#2}\big\}} {\big\{{#1},{#2}\big\}} {\{{#1},{#2}\}}{\{#1,#2\}}}}
 \newcommand{\Dvg}[2][]{(\hspace{-1pt}\partial_{#1}#2^{#1\!})\hspace{-1pt}}   %# divergence of the vector
 \newcommand{\LieB}[4][]{{\mathchoice{{\big[{#3},{#4}\big]^{#2}}}
                         {{\big[{#3},{#4}\big]^{#2}}}
                         {{[{#3},{#4}]^{#2}}}
                         {{[{#3},{#4}]^{#2}}} }}
 \newcommand{\CT}{\mathcal{T}}
 \newcommand{\ader}{\hspace{1pt}\accentset{\leftrightarrow}{\partial}\hspace{-1pt}} %% alternating derivative
  \bea{GUMG_Algebra_offshell} %From GUMGr_NDV file \ref{PB_GUMGr_B1}
   \begin{array}{lll}
     \PB{\sint {\gxi}^\zn\Hs_\zn}{\sint {\geta}^\zm\Hs_\zm}
     \!\!& = &\!\!
      \sint  (\gxi^\zm \partial_\zm \geta^\zn {-} \geta^\zm \partial_\zm \gxi^\zn) %\LieB{\zn}{\gxi}{\geta}
     \Hs_\zn
     , \vphantom{\big|^I}
     \\
     \PB{\sint {\gxi}^\zn\Hs_\zn}{\sint \geta^{\zm}\CT_{,\zm}}
     \!\!& = &\!\!
    % - \sint {\gxi}^\zn \Dvg{\geta}_{,\zn}\CT + \sint \Dvg{\gxi} \Dvg{\geta} \big( \WW+\TDer{\ln\WW}{\ln \sqrg} \big) \CT
    % \;=\;
     \sint {\gxi}^\zn \Dvg{\geta}\, \CT_{,\zn} + \sint \Dvg{\gxi} \hspace{1pt} \Dvg{\geta} \,\OOmega\, \CT
     , \vphantom{\big|^I}
     \\
     \PB{\sint {\gxi}^{\zn}\CT_{,\zn}}{\sint {\geta}^{\zm}\CT_{,\zm}}
     \!\!& = &\!\!
     \sint \big(\Dvg{\gxi} \,\ader_{\zn} {\Dvg{\geta}}\big) \g^{\zn\zm} \FF^2 \WW^2 \Hs_\zm
     ,  \vphantom{\big|^I}
     \\
   \end{array}\;
  \eea
and
  %\vspace{-8mm}
   \bea{parGUMG_Algebra_offshell}
   \begin{array}{lll}
    \PB{\sint \geps (\ptf_\io \hspace{-1pt}{+} \FF\Hl)}{\sint \geps' (\ptf_\io \hspace{-1pt}{+} \FF\Hl)}
    \hspace{-10mm}\!&& \hspace{5mm} =
    %0
    %+
    \sint
    %\big(\geps \partial_{\zn} \geps' {-} \geps' \partial_{\zn} \geps\big)
    \big(\geps \ader_{\zn} \geps'\big)
    \,\FF^2 \g^{\zn\zk} \Hs_\zk
   , \vphantom{\big|^I}
   \\
     \PB{\sint \geps (\ptf_\io \hspace{-1pt}{+} \FF\Hl)}{\sint \geta^\zn\Hs_\zn}
   \!\!\!& = &\!\!
   - \sint \geps_{,\zn} \geta^\zn \HIDE{\WW^{-1}} {\tfrac1\WW\,} \CT
          + \sint \geps \Dvg{\geta} \,\CT
         % + 0
   \!, \;\;\vphantom{\big|^I}
   \\
     \PB{\sint \geps (\ptf_\io \hspace{-1pt}{+} \FF\Hl)}{\sint \geta^\zn \CT_{,\zn}}
   \!\!\!& = &\!\!
   %- \sint \big(\geps \partial_{\zn} \Dvg{\geta}\WW {-} \Dvg{\geta}\WW\partial_{\zn} \geps\big)
   - \sint \big(\geps \ader_{\zn} \Dvg{\geta}\WW \big)
   \FF^2 \g^{\zn\zk} \Hs_\zk
   \vphantom{\big|^I}
   \\
   && \hspace{5mm}
   - \sint \geps \Dvg{\geta} \,\TTheta \,\trpg\,\CT
   , \vphantom{\big|^I}
   \\
   \end{array}
   \;\;\,
 \eea
where we use shorthand notations $\CT \defeq \WW\FF\Hl$, $\Dvg{\gxi} \defeq \partial_\zn\gxi^\zn$, $(f \ader_\zn g) \defeq (f \partial_\zn g {-} g \,\partial_\zn f)$ and introduce
 $ %  \bea{def_OOmega_TTheta}
    \OOmega(\argsqrg) \defeq  \TDer{\ln{\WW}}{\ln{\!\sqrg}} {\,+\,} \WW {\,+\,} 1\,,
 $
 \;% \qquad
 $   \TTheta(\argsqrg) \defeq  \tfrac{1}{\Ddim{-}2} \TDer{\ln{\WW}}{\ln{\!\sqrg}}\,\tfrac{\FF}{\sqrg}
    \,,
    %\nonumber\\
 $ %  \eea
it follows that the Poisson bracket matrix for the constraint set $\big(\ptf_\io{+}\FF\Hl \ssep \Hs_\zn \ssep (\WW\FF\Hl)_{,\zm}\big)$ has a greater rank than for the initial set $\big(\Hs_\zn \ssep (\WW\FF\Hl)_{,\zm}\big)$ \footnote{
 Note that in the right-hand side of (\ref{GUMG_Algebra_offshell}) and (\ref{parGUMG_Algebra_offshell}), $\CT \defeq \WW\FF\Hl$ (or,  more precisely, its average component) does not vanish on shell. The  exceptional cases with $\OOmega \tequiv 0$ correspond either to {\UMG}, $\WW \tequiv \wwc \teq {-} 1$, or to models with pathological constraint structures, and are thus excluded from regular {\GUMG} \cite{AltActionFull:2024}.
}.
To maintain the rank and satisfy the correspondence condition, one may extend all constraints with $O(\tf^\io_{,\zn})$ terms.
 %% While a formal solution can be formally constructed,
 %% it is generally impractical\HIDE{ for most of field theories,} as
This extension results in an infinite series in powers of $\tf^\io$  and its spatial derivatives. Furthermore, since the structure functions %%on the right-hand side of
in (\ref{GUMG_Algebra_offshell}),(\ref{parGUMG_Algebra_offshell})
are spatial differential operators, and some constraints are functionally incomplete, the result becomes uncontrollably nonlocal.

\if{
At the complete constraint surface of the theory ($\Hs_\zn=0$, $\CT_{,\zm}\!=0$) one gets
  \beq{GUMG_Algebra_onshell} %From GUMGr_NDV file \ref{PB_GUMGr_B1}
     \begin{array}{lll}
       \PB{\sint {\gxi}^\zn\Hs_\zn}{\sint {\geta}^\zm\Hs_\zm}
       & \weq & 0
       \, , \vphantom{\big|^I}\\
       \PB{\sint {\gxi}^\zn\Hs_\zn}{\sint \geta^{\zm}\CT_{,\zm}}
       & \weq & \sint \Dvg{\gxi} \Dvg{\geta} \,\OOmega\, \CT
       \, , \vphantom{\big|^I}\\
     %  \PB{\sint \gxi^{\zn}\CT_{,\zn}}{\sint {\geta}^\zm\Hs_\zm}
     %  & \weq & - \sint \Dvg{\gxi} \Dvg{\geta} \,\OOmega\, \CT \, , \\
       \PB{\sint {\gxi}^{\zn}\CT_{,\zn}}{\sint {\geta}^{\zm}\CT_{,\zm}}
       & \weq & 0
       \, , \vphantom{\big|^I} \\
     \end{array}
  \eeq
which shows that rank of Poisson bracket matrix is \emph{nonconstant} on the constraint surface.
}\fi

\vspace{1mm}

However, for field theories in which the \NBT{weighted} average-free part of the Hamiltonian density\HIDE{ vanishes on the complete constraint surface of} is constrained in the original theory, the equivalent parameterized action with local auxiliary fields can be constructed in a concise explicit form at the cost of simple spatial nonlocality.
 %\footnote{ For the restricted theories \cite{Barvinsky:2022guw}, where Hamiltonian usually contains the remnants of the restricted constraint structures from the parental theory with nonabelian algebras, the partial reincarnation of (a part of) the lost parental constraints via secondary constraints in restricted theory is a typical case \cite{Nesterov:Restricted}.}
   %%We need parameterized action to gather full Hamiltonian density term in a functionally complete constraint, and this can be achieved.
 %
Conceptually, the desired representation of the theory is achieved in two steps,  each of which clearly preserves the physical equivalence of the representations.
 %% The price for this is a possible spatial non-locality, which gives rise to delocalization operator $\op{E}$ in (\ref{ActionHTlike_GUMG},\ref{ActionHTlike_GUMG_ccf0}).
 %%, which can be effectively transferred to the auxiliary sector of the fields.
 %
 %%For generic {\GUMG} theory this works as follows.
First, a universally valid homogeneous time parameterization
 \footnote{Homogeneous parameterization in field theory is generally possible because the dynamical and constraint structures in the\HIDE{ extended} canonical theory (\ref{extAction_GUMG}) are harmonized by the\HIDE{ underlying} Dirac consistency procedure\HIDE{ \cite{Henneaux:1992ig}}: the spatially integrated Hamiltonian density, even when shifted by a commuting element, remains in involution with the complete set of the original constraints.}
is applied to (\ref{extAction_GUMG}) by introducing homogeneous fields $\hmg{\tf}^\io(\taux), \hmg{\ptf}_\io(\taux), \hmg{\replm}^\io(\taux)$. Then, the average-free parts of these fields are added via\HIDE{ initially decoupled} separate pure-gauge combination
   $
     \tauxdot{\inh{\tf}}^\io \inh{\ptf}_\io  {\,-\,} \inh{\replm}^\io \,\inh{\ptf}_\io
   $.
After this, it becomes possible to combine a functionally complete $\Hl$  term in a constraint, which is necessary to obtain the desired generalization of the  Henneaux-Teitelboim representation.
% When the weighting factor has spatial dependence, this combination will generate a {simple} spatial nonlocality\HIDE{ of a simple type}, which due to linearity in $\Hl$ necessarily would be a nondegenerate operator multiplier.
% Though the latter can be attached to the auxiliary Lagrange multiplier field, it will manifest itself in the dynamics of the original fields, since the nonlocality just reveals the actual dynamic property hidden in the original\HIDE{ non-parameterized} formulation.
   %% Although the nonlocality can be attached to the auxiliary fields, it manifests itself in the dynamics of the original fields, being their\HIDE{ actual} dynamic property hidden in the original\HIDE{ non-parameterized} formulation.

For the general {\GUMG} model, this parameterization yields the equivalent parameterized action in the form
 \bea{paramAction_min_GUMG} %% (\ref{paramAction_min_GUMG})
    S_{par}\HIDE{^{\,min}} [\g,\pg,\tf^\io\!,\ptf_\io,\replm^\io\!,\repNs,\replmCT]
    &=&\! \int\! d\taux \hspace{1pt} \dsx \,
    \LiB
     \tauxdot{\g}_{\zm\zn}\pg^{\zm\zn}
     \!+ \tauxdot{\tf}^\io\ptf_\io
     \nonumber\\
     &&
     \hspace{-85pt}
     - \replm^\io ( \ptf_\io {+} \hmg{\FF \Hl} )
     - \repNs^\zn \Hs_\zn
     - \replmCT^\zm (\WW\FF\Hl)_{,\zm}
     \RiB
     \,,
   \qquad% \nonumber\\
  \eea
where auxiliary fields are combined into\HIDE{ the functionally full} local fields ${\tf}^\io(\taux,\sx), \ptf_\io(\taux,\sx), \replm^\io(\taux,\sx) $. It can be verified\HIDE{ \cite{AltActionFull:2024}} that the rank of the matrix of Poisson brackets of constraints does not change\HIDE{ on the constraint surface}, the  gauge ${\tf}^\io{-\,}\taux \teq 0$ is accessible, and the reduction  with respect to\HIDE{ the second-class constraint pair} $\big(\ptf_\io {+} \hmg{\FF \Hl} \ssep {\tf}^\io{-}\taux \big)$  reproduce (\ref{extAction_GUMG})
 \footnote{The rank property may be checked using (\ref{parGUMG_Algebra_offshell}), whose right-hand side vanishes on the constraint surface after substitution $\geps\to\hmg{\geps}$, $\geps'\to\hmg{\geps'}$, caused by the change of the parameterized Hamiltonian constraint. The correspondence property is evident since the canonical action of $\ptf_\io {+} \hmg{\FF \Hl}$ is obviously\HIDE{ completely} transversal to $\tf^\io\!{-\,}\taux$.}.
The form of the new first-class constraint\HIDE{ at non-GR branch}, introduced during such parameterization, can be guessed from general {principles}. This is
 $ % \beq{I-class_constr}
  \CPI \teq
  %\sint \geps (
   \ptf_\io {+\,} \hmg{\FF\Hl} {+\,} \hmg{U_0^\zn \Hs_\zn}
  %  )
  \,, % \\
 $ % \eeq
where $U_0^\zn(\g,\pg)$ is an\HIDE{ particular} on-shell solution for the Lagrange multipliers $\Ns^\zn$, which in {\wGUMG} can be chosen equal to\HIDE{ as} zero,
 but is nontrivial in general {\GUMG}
 \footnote{The combination ${\FF\Hl} {+\,} {U_0^\zn \Hs_\zn}$, which appear in $\CPI$, is the so-called first-class Hamiltonian density \cite{Henneaux:1992ig}.
 Also we recall that the discussed properties of the constraints are related to the physical space of the theory outside the GR branch \cite{Barvinsky:2019qzx}.
}.
The new first-class constraint $\sint \geps \CPI$ and the transverse\HIDE{ spatial} diffeomorphisms $\sint \gzetat^\zn \Hs_\zn$, ($\partial_\zk\gzetat^\zk \teq 0$), generate the canonical gauge algebra of the model. The secondary constraint and longitudinal spatial diffeomorphisms are second-class.

For {\wGUMG} ($\wwc{\,\neq\,}{-}1$), this reproduces the constraint structure derived\HIDE{ earlier} from the local action (\ref{paramAction_naive_AMG}).
The newly introduced parameterization symmetry\HIDE{ of (\ref{paramAction_min_GUMG})} acts homogeneously on the metric phase-space fields, with the reason for this being more transparent in this scheme. Only the average component of ${\FF\Hl}$ from $\CPI$ generate gauge transformations, while its average-free part\HIDE{ is efficiently constrained by the secondary constraint and} does not.

For the {\UMG} case $\WW\tequiv\HIDE{\wwc\teq}{-}1$, parameterization (\ref{paramAction_min_GUMG}) clarifies how the lacking homogeneous mode of $\sqrg^{-1}\Hl$ enter the game, with the inhomogeneous part of the new symmetry acting entirely within the auxiliary sector.
Unlike in {\GUMG}, the inhomogeneous part of the Hamiltonian constraint and the longitudinal part of spatial diffeomorphisms are initially first-class, so the homogeneous mode just completes the original gauge symmetry to full diffeomorphism invariance in the metric sector.

This parameterization scheme also clarifies that in {\GUMG} only the homogeneous time-reparameterization transformations are shared by both the metric and the auxiliary fields, while the remaining average-free symmetries act independently in these sectors.

%\vspace{2mm}

To derive the generalized Henneaux-Teitelboim form of the action from (\ref{paramAction_min_GUMG}), we rearrange its constraint basis\HIDE{,
particularly to separate the contributions of the {\GR} canonical terms}.
The constraint equations $\ptf_\io {+} \hmg{\FF \Hl} \teq 0$, $ (\WW \FF \Hl)_{,\zm} \teq 0$
 are equivalent to the set $\ptf_\io {+} \hmg{\FF \Hl} {+} \WW^{-1} (\inh{\WW \FF \Hl}) \teq 0$,\, $(\WW\FF\Hl)_{,\zm} \teq 0$.
The first constraint in the latter set can be recast as $\ptf_\io {+} \op{E}\FF \Hl \teq 0$, where the operator $\op{E}$, with kernel (\ref{opE_kernel}), is nondegenerate and acts to the right on the Hamiltonian density $\FF \Hl$.
This is the origin of the spatial nonlocality in the parameterized {\GUMG}\HIDE{theories with nonconstant $\WW$}: it is generally not possible to merge functionally incomplete constraints into functionally complete local combinations.

To assemble the dynamical metric dependence into a\HIDE{ desired} {\GR}-like combination, the\HIDE{ functionally full} $\Hl$ must be consolidated into a single constraint.
 % This can be achieved by virtue of nondegeneracy of $\op{E}$.
Substituting $\FF\Hl \teq {-}\op{E}^{-1}\ptf_\io$, where $\op{E}^{-1}$ is defined in (\ref{invE_kernel}), into the secondary constraint results in $ (\WW\op{E}^{-1}\ptf_\io )_{,\zm} \teq 0$. The latter\HIDE{ constraint} looks awkward but simplifies to $\ptf_{\io,\zm} \teq 0$ \cite{AltActionFull:2024}. In the new constraint basis the canonical action becomes
\vspace{-1mm}
   \bea{paramAction'_GUMG} %(\ref{parAction_BLRG})
    \!{S}_{par'}[\g,\pg,\tf^\io\!,\ptf_\io,\replm,\repNs,\lmptf]
    &=& %\!\!&=& \!\!\!
    \!\int \! d\taux \hspace{1pt} \dsx \, \LiB
     \tauxdot{\g}_{\zm\zn} \pg^{\zm\zn}
     \!+
     \tauxdot{\tf}^\io \ptf_\io
     \nonumber\\
     &&
     \hspace{-60pt}
     - \replm ( \ptf_{\io} {+} \op{E}{\FF \Hl} )
     - \repNs^\zn \Hs_\zn
     - \lmptf^\zm {\ptf_\io}_{,\zm}
     \RiB
     . \;\;
 \eea
The equivalence of (\ref{paramAction_min_GUMG}) and (\ref{paramAction'_GUMG})
is straightforward, as they are related by a nondegenerate redefinition of the Lagrange multipliers,
$
   \replm^\io
   \teq
    \replm {\,-\,} \sDvg[\zk]{\lmptf} \,, %%  - \sDvg[\zk]{\lmptf} \to
   \; \sDvg[\zk]{\replmCT} %\equiv \Bra{\inh{\lmv}}
   \teq
   {-} \inh{ \replm  \WW^{-1}\HIDE{\ofA} }  %% \ofA\to\WW^{-1}
   % \,.
$.

\vspace{0.6mm}

Finally, the effective lapse function $\repNl$ is recovered via the redefinition $\replm \teq \repNl \FF^{-1} \op{E}^{-1}$, which completes the formation of the canonical expression for the Einstein-Hilbert Lagrangian.
The Legendre reduction to Lagrangian action proceeds identically to {\GR},
which leads to the action (\ref{ActionHTlike_GUMG_ccf0}),
 \if{
  \bea{ActionHTlike_GUMG_ccf0}
    S_{_{{\GUMG}}}^{\ialt_0}[\Gaux,\CCof,\HTf]
    &=& \!\! \int \!d\taux\, \dsx \, \sqrt{|\Gaux|} \Big( \stR(\Gaux) - \sqrg^{-1}\FF^{-1}\op{E}^{-1}\CCof \Big)
    +  \!\int \!d\taux\, \dsx \, \partial_\Zm \HTf^\Zm  \CCof
    \;, \quad
   % \nonumber
  \eea
 }\fi
where spacetime metric is
 $ % \beq{Gaux_ADM_components}
  \Gaux_{\Zm\Zn}d{\Xaux}^{\Zm}d{\Xaux}^{\Zn}\!
   \teq\,  \g_{\zm\zn} (d\sx^\zm {+} \repNs^\zm d\taux)(d\sx^\zn {+} \repNs^\zn d\taux) - \repNl^2 d\taux d\taux ,
   \vphantom{\hat{I}}
 $ % \eeq
the cosmological-constant field $\CCof$ is just the renamed $\ptf_\io$, and the Henneaux-Teitelboim field is \,$\HTf^\Zm\HIDE{ \equiv \big(\HTf^0, \HTf^\zn \big)} \teq \big(\tf^\io, \lmptf^\zm \big) $.

%%-------------------------=%        ******         %=-------------------------%%
%%-------------------------=%        ******         %=-------------------------%%
%%-------------------------=%        ******         %=-------------------------%%

\vspace{-0.7mm}

    \section{Dynamical properties}
  %\hspace{\parindent}
  %
The variational equations of motion for action $S_{\ialt}[\Gaux,\CCf,\HTf]$ (\ref{ActionHTlike_GUMG}) with respect to the auxiliary fields  $\HTf^\Zn$, \HIDE{which enter the action linearly,}
  \beq{EoM_tef_GUMG}
  %  \VDer{S_{\ialt}}{\HTf^\Zn} \eomeq  0
  %  \qquad\Rightarrow\qquad
    \partial_\Zn (  \op{E}  \FF  \sqrg \CCf )
    \,= \,
     \partial_\Zn  \CCof
    \,\eomeq \,
     0
    \, ,
  \eeq
generalize\HIDE{ {\UMG}} the cosmological constant behavior, implying that  $\CCof \HIDE{= \ptf_\io}$ is an unfixed on-shell spacetime constant.
Operator $\op{E}$ (\ref{opE_kernel}) introduces spatial nonlocality into the classical behavior of the cosmological-constant field $\CCf$,
  \beq{ccf_onshell_GUMG}
    \CCf
    \,\eomeq  \sqrg^{-1} \FF^{-1} \op{E}^{-1} \cco\,
    = \sqrg^{-1} \FF^{-1} \frac{\WW^{-1}}{\,\hmg{\WW^{-1}\!}\,\vphantom{I^{|^I}}} \, \cco
  \,,
  \eeq
where $\cco$ is an\HIDE{ unfixed} on-shell constant value of $\CCof$ and $\hmg{\WW^{-1}}(\taux)$ is the\HIDE{ spatial} average of  $\WW^{-1}(\sqrg {\scalebox{0.8}{\mbox{\ensuremath{(\taux,\sx)}}}})$
 over $\taux \teq const$  hypersurface.
For \wGUMG\ subfamily (\ref{AMG_part_case})  $ \WW^{-1} \!/ {\raisebox{-1.5pt} {$\ensuremath \,\hmg{\WW^{-1}\!} $}}\, \teq 1$ and the spatial nonlocality disappears.
In the unimodular case (\ref{UMG_part_case}),  $\CCf$ coincides with $\CCof$ and on shell equals to $\cco$.

The variational equation with respect to $\CCf$ gives
 \beq{EoM_ccf_GUMG}
  %  \VDer{S_{\ialt}}{\CCf}
  % \eomeq  0
  % \qquad\Rightarrow\qquad
    \sqrt{|\Gaux|}
    \,\eomeq\,  \partial_\Zm \HTf^\Zm  \op{E}  \FF  \sqrg
   % \,\eomeq\, \tauxdot{\hmg{\tf}}\,  \FF  \sqrg
    \,,
 \eeq
with $\op{E}$ acting to the left.

For {\GUMG} models, unlike the exceptional unimodular case (\ref{UMG_part_case}), two \HIDE{spatially-}average-free second-class constraints arise\HIDE{\footnote{On the non-GR branch.}}.
Differential consequences of the equations of motion impose two\HIDE{ additional} constraints on metric components, which in canonical treatment are the Lagrange multipliers at these constraints. First, the lapse function $\repNl $ satisfies
 \beq{inhrepNl_onshell}
   \inh{\repNl \FF^{-1}}
   \eomeq\,
    0
   \,.
  \eeq
This\HIDE{ relation} and (\ref{EoM_ccf_GUMG}) imply $\!\inh {\partial_\Zm \HTf^\Zm  \op{E}} {\,\sim\, } ( \inh {\partial_\Zm \HTf^\Zm  \WW^{-1} } \! ) \!\eomeq 0$, leading to
  \beq{Dvg_HTf_on_shell}
    \partial_\Zm \HTf^\Zm  \op{E}
    \,\eomeq\, \hmg{\partial_\Zm \HTf^\Zm}
    =\, \tauxdot{\hmg{\tf}}^\io
    \,,
  \eeq
and reinstating the\HIDE{ initial} {\GUMG} restriction condition (\ref{GUMG_restriction}) as
 $ % \beq{repNl_onshell}
   \repNl
   {\,\eomeq\,\,}  \tauxdot{\hmg{\tf}}^\io \FF(\argsqrg)
   \,.
 $ % \eeq
Factor
$\tauxdot{\hmg{\tf}}^\io(\taux)$ reflects time reparametrization gauge ambiguity in  parameterized {\GUMG}\HIDE{ setup}\HIDE{ (\ref{ActionHTlike_GUMG})}, which acts on the gravitational variables homogeneously, and is equal to $1$ in the correspondence gauge $\tf^\io {-\,} \taux \teq 0$.
\HIDE{With (\ref{Dvg_HTf_on_shell}) }The second constraint concerns Lagrange multipliers for\HIDE{ second-class} longitudinal spatial diffeomorphisms,
 \beq{DvgrepNs_onshell}
   \begin{array}{lcl}
   \partial_\zn\repNs^\zn
    \eomeq
    \tauxdot{\hmg{\tf}}^\io\,
     \HIDE{\,\partial_\zn U_0^\zn \;=\;}
    \big( \OOmega^{-1}\TTheta \trpg
      - \OOmega^{-1}
      {\sint \OOmega^{-1}\TTheta \trpg}\big{/}{\sint \OOmega^{-1}}
      \big)
   \,,\\
   \end{array}
 \eeq
where
 $ %  \bea{def_OOmega_TTheta}
    \OOmega(\argsqrg) \defeq  \TDer{\ln{\WW}}{\ln\sqrg} + \WW + 1\,,
 $
 \,% \qquad
 $   \TTheta(\argsqrg) \defeq  \tfrac{1}{\Ddim{-}2} \TDer{\ln{\WW}}{\ln{\sqrg}}\,\tfrac{\FF}{\sqrg}
    \,.
    %\nonumber\\
 $ %  \eea

 % - - - - -

Finally, the {variational derivative} of\HIDE{ the action} $S_{\ialt}[\Gaux,\CCf,\HTf]$ %%\HIDE, (\ref{ActionHTlike_GUMG}),
with respect to\HIDE{ the spacetime metric} $\Gaux_{\Zm\Zn}$
%%acquires the form of the Einstein equations\HIDE{ with matter}
leads to the Einstein equations
%\vspace{-1mm}
 \beq{EoM_GG_AMG}
 %  \VDer{S_{\ialt}}{\Gaux_{\Zm\Zn}}
 % \eomeq 0
 % \qquad\Rightarrow\qquad
    %\sqrt{|\Gaux|} \,
    \stR^{\Zm\Zn} \!- \tfrac12 \stR \,\Gaux^{\Zm\Zn} % + \tfrac12\Lambda\GG^{\Zm\Zn})
    \eomeq\,
    %\sqrt{|\Gaux|} \,
    \tfrac12 \repTSEt^{\Zm\Zn} %%_{\scriptscriptstyle\! \GUMG}
    \,,
 \eeq
where the Einstein tensor for spacetime metric $\Gaux_{\Zm\Zn}$ is on the left, and terms proportional to $\CCf$ are attributed to the matter sector,
 %% $S_{mat}=\!\int \!d\taux\, \dsx \, (-\sqrt{|\Gaux|}\CCf + \partial_\Zm \HTf^\Zm  \op{E}  \FF  \sqrg \CCf)$,
whose  contribution is encoded on the right-hand side in the energy-momentum tensor
   $ \repTSEt^{\Zm\Zn} \defeq  \frac{2}{\sqrt{|\Gaux|}} \VDer{S_{mat}}{\Gaux_{\Zm\Zn}}$.
The latter can be cast on shell to
 \beq{Stress-Energy_Tensor_GUMG}
    \repTSEt^{\Zm\Zn}
   \!\eomeq
    \CCf \, n^{\Zm}n^{\Zn}
    + \WW
    \CCf  \big(\Gaux^{\Zm\Zn}
    \!+ n^{\Zm}n^{\Zn}\big)
    \,,
 \eeq
where $n^\Zm$ is a unit vector field orthogonal to constant-time hypersurfaces. In (\ref{Stress-Energy_Tensor_GUMG}) $\,\repeSEt \eomeq \CCf \,$ is the energy density and $\,\reppSEt \eomeq \WW\CCf\,$ is the pressure, reproducing the perfect fluid equation of state (\ref{PerfectFluid_Eq_of_state}),
 \beq{}
   \reppSEt \,\eomeq\, \WW \repeSEt
  \,.
 \eeq
 %%with dynamic parameter $\WW=\WW(\argsqrg)$.
 Note that for nonconstant $\WW(\argsqrg)$ the on-shell behaviour of energy and pressure bears the imprint of the spatial nonlocality due to (\ref{ccf_onshell_GUMG}).

%%-------------------------=%        ******         %=-------------------------%%
%%-------------------------=%        ******         %=-------------------------%%
%%-------------------------=%        ******         %=-------------------------%%

\vspace{-0.7mm}

\section{Conclusions}

%% v3
%% In this letter we presented an alternative formulation of generalized unimodular gravity. It can be formulated by the action, similar to Henneaux-Teitelboim action for unimodular gravity. We briefly discussed its derivation, including issues of consistent parameterization. Also we highlighted basic dynamical features of the model in alternative formulation, including the appearance of the spatial nonlocal structure, and briefly mentioned the structure of the canonical gauge algebra.
%% The line of reasoning and underlying techniques can be generalized to restricted gravity models with local restrictions of more general types, fixing lapse function in terms of quantities, built from the induced spatial metric.

In this letter, we introduced an alternative formulation of generalized unimodular gravity (\GUMG), inspired by the Henneaux-Teitelboim action for unimodular gravity. We derived this formulation, addressing issues of consistent parameterization, and highlighted its key dynamical and gauge features, including the emergence of spatial nonlocality.
A detailed discussion, including a complete analysis of the gauge structure of the Lagrangian theory, a comprehensive specification of exceptional particular cases, various conceptual and calculational subtle aspects of the formalism\HIDE{, and a note on the quantum implications of the nonlocality,} are presented in \cite{AltActionFull:2024}.

The parameterization approach and the alternative action in the generalized Henneaux-Teitelboim form, along with the underlying techniques for dealing with functionally incomplete constraints, can be extended to restricted gravity models with more general restriction conditions. Also these methods and some results can be directly applied to other field theories where the gradient of the weighted Hamiltonian density vanishes on the constraint surface, which may be generated by analogous restriction of a Lagrange multiplier at Hamiltonian constraint in diffeomorphism-invariant theories.
%Such theories can be generated by restricting diffeomorphism-invariant gauge theories, where the restriction relates the field serving as the Lagrange multiplier in the Hamiltonian constraint (analogous to $\Nl$) to a rather arbitrary function of the Lagrangian dynamical fields (analogous to $\g_{\zm\zn}$).

In conclusion, we remark on the nature of the emerged spatial nonlocality. While one might view the\HIDE{ spatial} nonlocality encoded in $\op{E}$ as an artifact of the ``unfortunate'' parameterization by local auxiliary fields, we argue that this nonlocality is implicitly present in the physical sector of \,$\WW {\,\neq\,\,} const$\, {\GUMG} models already in the original formulation.
In the canonical framework, the integrated Hamiltonian density $\hmg{\FF\Hl}$ is a gauge-invariant constant of motion, with its on-shell values parameterized by a constant $-\cco$. The secondary constraint $\inh{\WW\FF\Hl}$ is a gauge invariant that vanishes on shell. Thus, $\op{E}\FF\Hl$ is functionally complete (but spatially nonlocal) gauge-invariant constant of motion --- a property that the alternative parameterized representation makes explicit via a constraint.

The significance of this observation is further supported by the perspective of the  restricted-theory formalism \cite{Barvinsky:2022guw,DNRestricted:2025}, which sheds light on the connection between the\HIDE{ one-loop} effective actions of the parental and restricted gauge theories. The important point is to understand and conveniently describe the difference between the physical spaces of the parental and restricted theories, when the restriction is not a pure gauge fixing \cite{Barvinsky:2022guw}.
In {\GUMG} the deformation of the physical space is parameterized by a constant. In particular, the physical trajectories of the parental theory (\GR) are tied to the constraint surface $\Hl \teq 0$,\, $\Hs_\zn \teq 0$, in the restricted theory (\GUMG) the constraint surface expands
 % \footnote{The corresponding contraction of the physical space \cite{Barvinsky:2022guw} is not confined to the\HIDE{ concise} phase space, as it restricts a global\HIDE{ gauge-invariant} mode of the quantity involving the Lagrange multiplier $\Nl$. However, it can also be  inferred from the actions of the alternative representation\HIDE{ (\ref{ActionHTlike_GUMG})}.}
to
  $ \Hl
    \teq {-}\FF^{-1}\op{E}^{-1} \cco
    \,\teq {-}\FF^{-1} \frac{\WW^{-1}}{\,\hmg{\!\WW^{-1}\!}\,\vphantom{I^{|^I}}} \, \cco
  $,
\, $\Hs_\zn \teq 0$.

We expect that the results and methods developed for generalized unimodular gravity will be useful for further study of similar theories and their quantum properties.

\begin{acknowledgments}
We wish to thank Andrei Barvinsky\HIDE{ and other colleagues} for stimulating discussions. The research was supported by the Russian Science Foundation grant No \href{https://rscf.ru/en/project/23-12-00051/}{23-12-00051}.
% put your acknowledgments here.
\end{acknowledgments}

% Create the reference section using BibTeX:
\bibliographystyle{ieeetr}%{plainnat}%{ieeetr}%{unsrt}%{ieeetr}%{siam}%{plain} {plainnat} {apalike} {IEEEtran}
\bibliography{grav}%{basename of .bib file}

\begin{thebibliography}{10}

\bibitem{Einstein1919UMG}
A.~Einstein, ``{Do gravitational fields play an essential part in the structure
  of the elementary particles of matter?},'' {\em Sitzungsber. Preuss. Akad.
  Wiss. Berlin (Math. Phys.)}, vol.~1919, pp.~349--356, 1919.
%\newblock Translated and included in The Principle of Relativity, by H.A.
%  Lorentz et al. (Dover Press, New York, 1923).

\bibitem{Weinberg:1988cp}
S.~Weinberg, ``{The Cosmological Constant Problem},'' {\em Rev. Mod. Phys.},
  vol.~61, pp.~1--23, 1989.

\bibitem{Unruh:1988in}
W.~G. Unruh, ``{A Unimodular Theory of Canonical Quantum Gravity},'' {\em Phys.
  Rev. D}, vol.~40, p.~1048, 1989.

\bibitem{Ng:1990xz}
Y.~J. Ng and H.~van Dam, ``{Unimodular Theory of Gravity and the Cosmological
  Constant},'' {\em J. Math. Phys.}, vol.~32, pp.~1337--1340, 1991.

\bibitem{Bousso:2007gp}
R.~Bousso, ``{TASI Lectures on the Cosmological Constant},'' {\em Gen. Rel.
  Grav.}, vol.~40, pp.~607--637, 2008.

\bibitem{Bufalo:2015wda}
R.~Bufalo, M.~Oksanen, and A.~Tureanu, ``{How unimodular gravity theories
  differ from general relativity at quantum level},'' {\em Eur. Phys. J. C},
  vol.~75, no.~10, p.~477, 2015.

\bibitem{Henneaux:1989zc}
M.~Henneaux and C.~Teitelboim, ``{The Cosmological Constant and General
  Covariance},'' {\em Phys. Lett. B}, vol.~222, pp.~195--199, 1989.

\bibitem{Barvinsky:2017pmm}
A.~O. Barvinsky and A.~Y. Kamenshchik, ``{Darkness without dark matter and
  energy -- generalized unimodular gravity},'' {\em Phys. Lett. B}, vol.~774,
  pp.~59--63, 2017,
 % doi:10.1016/j.physletb.2017.09.045,
  arXiv:\href{https://arxiv.org/abs/1705.09470}{1705.09470} [gr-qc].

\bibitem{Arnowitt:1959ah}
R.~L. Arnowitt, S.~Deser, and C.~W. Misner, ``{Dynamical Structure and
  Definition of Energy in General Relativity},'' {\em Phys. Rev.}, vol.~116,
  pp.~1322--1330, 1959.

\bibitem{Misner:1973prb}
C.~W. Misner, K.~S. Thorne, and J.~A. Wheeler, {\em {Gravitation}}.
\newblock San Francisco: W. H. Freeman, 1973.

\bibitem{Barvinsky:2019agh}
A.~O. Barvinsky, N.~Kolganov, A.~Kurov, and D.~Nesterov, ``{Dynamics of the
  generalized unimodular gravity theory},'' {\em Phys. Rev. D}, vol.~100,
  no.~2, p.~023542, 2019,
 % doi:10.1103/PhysRevD.100.023542,
  arXiv:\href{https://arxiv.org/abs/1903.09897}{1903.09897} [hep-th].

\bibitem{Barvinsky:2019qzx}
A.~O. Barvinsky and N.~Kolganov, ``{Inflation in generalized unimodular
  gravity},'' {\em Phys. Rev. D}, vol.~100, no.~12, p.~123510, 2019,
 % doi:10.1103/PhysRevD.100.123510,
  arXiv:\href{https://arxiv.org/abs/1908.05697}{1908.05697} [gr-qc].

\bibitem{AltActionFull:2024}
D.~Nesterov, ``{Alternative action for generalized unimodular gravity}.''
  submitted for publication, 2025,
  arXiv:\href{https://arxiv.org/abs/2505.13548}{2505.13548} [gr-qc].

\bibitem{Henneaux:1992ig}
M.~Henneaux and C.~Teitelboim, {\em {Quantization of gauge systems}}.
\newblock 1992.

\bibitem{Barvinsky:2022guw}
A.~O. Barvinsky and D.~V. Nesterov, ``{Restricted gauge theory formalism and
  unimodular gravity},'' {\em Phys. Rev. D}, vol.~108, no.~6, p.~065004, 2023,
 % doi:10.1103/PhysRevD.108.065004,
  arXiv:\href{https://arxiv.org/abs/2212.13539}{2212.13539} [hep-th].

\bibitem{DNRestricted:2025}
 in preparation.

\end{thebibliography}

\end{document}